\documentclass[prl,twocolumn,
superscriptaddress,
nofootinbib,
 amsmath,amssymb,
 aps,
floatfix
]{revtex4-2}

\usepackage{graphicx}
\usepackage{dcolumn}
\usepackage{bm}
\usepackage{aas_macros}
\usepackage{xcolor} 

\definecolor{green}{rgb}{0,0.4,0}

\usepackage{hyperref}
\hypersetup{linkcolor=red,citecolor=blue,urlcolor=cyan}

\newcommand{\be}{\begin{equation}}
\newcommand{\ee}{\end{equation}}
\newcommand{\noi}{\noindent}

\begin{document}

\title{Binary coalescences as sources of Ultra-High Energy Cosmic Rays} 

\author{Jonas P. Pereira}%
\email{jpereira@camk.edu.pl}
\affiliation{Núcleo de Astrofísica e Cosmologia (Cosmo-Ufes) \& Departamento de Física, Universidade Federal do Espírito Santo, Vitória, 29075-910, ES, Brazil}
\affiliation{Nicolaus Copernicus Astronomical Center, Polish Academy of Sciences, Bartycka 18, 00-716, Warsaw, Poland}

\author{Carlos H. Coimbra-Ara\'ujo}
\affiliation{Departamento de Engenharias e Exatas, Universidade Federal do Paran\'a, Pioneiro, 2153, 85950-000, Palotina, PR, Brazil}
\affiliation{Programa de Pós-Graduação em Física Aplicada, Universidade Federal da Integração Latino-Americana, 85867-670, Foz do Igua\c{c}u, PR, Brazil}

\author{Rita C. dos Anjos}
\affiliation{Departamento de Engenharias e Exatas, Universidade Federal do Paran\'a, Pioneiro, 2153, 85950-000, Palotina, PR, Brazil}
\affiliation{Programa de Pós-Graduação em Física Aplicada, Universidade Federal da Integração Latino-Americana, 85867-670, Foz do Igua\c{c}u, PR, Brazil}
\affiliation{Programa de Pós-Graduação em Física e Astronomia, Universidade Tecnológica Federal do Paraná, Jardim das Americas, 82590-300, Curitiba, PR, Brazil}
\affiliation{Programa de Pós-Graduação em Física, Universidade Estadual de Londrina, Rodovia Celso Garcia Cid, Pr 445 Km 380, Campus Universitário, 86057-970, Londrina, PR, Brazil}

\author{Jaziel G. Coelho}
\affiliation{Núcleo de Astrofísica e Cosmologia (Cosmo-Ufes) \& Departamento de Física, Universidade Federal do Espírito Santo, Vitória, 29075-910, ES, Brazil}
\affiliation{Divis\~ao de Astrof\'isica, Instituto Nacional de Pesquisas Espaciais, S\~ao Jos\'e dos Campos, 12227-010, SP, Brazil}

\date{\today}

\begin{abstract}
\noindent Binary coalescences are known sources of gravitational waves (GWs) and they encompass combinations of black holes (BHs) and neutron stars (NSs). Here we show that when BHs are embedded in magnetic fields ($B$s) larger than approximately $10^{10}$ G, charged particles colliding around their event horizons can easily have center-of-mass energies in the range of ultra-high energies ($\gtrsim 10^{18}$ eV) and become more likely to escape. Such B-embedding and high-energy particles can take place in BH-NS binaries, or even in BH-BH binaries with one of the BHs being charged (with charge-to-mass ratios as small as $10^{-5}$, which do not change GW waveforms) and having a residual accretion disk. Ultra-high center-of-mass energies for particle collisions arise for basically any rotation parameter of the BH when $B \gtrsim 10^{10}$ G, meaning that it should be a common aspect in binaries, especially in BH-NS ones given the natural presence of a $B$ onto the BH and charged particles due to the NS's magnetosphere. We estimate that the number of ultra-high center-of-mass collisions ranges from a few up to millions before the merger of binary compact systems. Thus, binary coalescences may also be efficient sources of ultra-high energy cosmic rays (UHECRs) and constraints to NS/BH parameters would be possible if UHECRs are detected along with GWs. 
\end{abstract}

\maketitle

\noi {\bf {\it Introduction.---}} In 2015, the first direct detection of gravitational waves (GWs) from a binary black hole (BBH) merger by the Laser Interferometer Gravitational-Wave Observatory (LIGO) and the Virgo Collaboration, GW150914 \citep{abbott2016}, inaugurated the field of gravitational-wave astronomy. Shortly after, GW170817 \citep{abbott2017a}, a binary neutron star (BNS) coalescence presenting an electromagnetic signal across all electromagnetic spectrum \citep{abbott2017b}, firmly established the field of multimessenger astronomy. Now, GW events are ``common'' and so far around 90 of them--mostly BBH mergers \citep{abbott2021a,abbott2021b} but also BNS mergers [see \cite{spera} for a review]--have already been detected.

Third-generation GW detectors, such as the Cosmic Explorer \citep{2019BAAS...51g..35R} and Einstein telescope \citep{2020JCAP...03..050M} (and even 2.5-ones such as NEMO \citep{2020PASA...37...47A}) promise to significantly lower the uncertainties of GW observables and increase the number of detections by means of substantial improvements in sensitivity. We expect to detect around $10^5-10^6$ BBH (or BH-NS) events and $10^4-10^5$ BNS events every year with a single 3G GW detector \citep{2020JCAP...03..050M}. With multimessenger astronomy, follow-ups and simultaneous observations of GW events will become routine with electromagnetic, neutrino, and high-energy particle detectors. Each one of these messengers provides different windows onto BHs, NSs, quasars, blazars, supernovae, and other sources, and the complementarity of those observations is starting to have a deep impact on physics and astronomy. That will lead to an unprecedented richness of astrophysical data and, with it, the opportunity will present itself to probe many astrophysical and cosmological models, as well as 
many physical mechanisms.

One of extreme relevance concerns the understanding of the most powerful accelerators in the universe, which give rise to ultra-high energy cosmic rays (UHECRs) \citep{meszaros,montaruli,kotera2016}. Regarding the experimental picture of high-energy cosmic rays, it is well known that they interact with the atmosphere and produce air showers that can be detected by ground-based water Cherenkov detectors, scintillator surface detectors, or underground muon detectors. In particular, extensive air shower arrays, such as the Pierre Auger Observatory \citep{auger} or the Telescope Array \citep{ta}, already detect cosmic rays from $\sim 10^{16}$ eV to $\sim 10^{19}$ eV. The Southern Wide-field Gamma-ray Observatory (SWGO) would, for instance, measure in the next years the cosmic ray spectrum up to the so-called ``knee'', at $10^{15}$ eV \citep{swgo}. Other experiments such as H.E.S.S. \citep{hess}, VERITAS \citep{veritas}, FERMI-LAT \citep{fermi}, MAGIC \citep{magic} and LHAASO \citep{lhaaso} -- and in the future CTA \citep{cta}, among others -- are also able to constrain upper limit energies of the more energetic cosmic rays using gamma-ray observations, since they are possible outcomes of the cosmic ray propagation which contribute to the total flux measured from the source \citep{Supanitsky_2013, Anjos_2014, PhysRevD.96.023008,2021JCAP...10..023D,2022JCAP...10..041C}.

Cosmic rays encompass the most energetic particles detected by ground experiments, whose energies are up to ten million times those reached by the Large Hadron Collider. In particular, UHECRs propagate in the universe with energies beyond the so-called Greisen-Zatsepin-Kuzmin (GZK) limit ($E > 10^{18}$ eV) \citep{1966JETPL...4...78Z, PhysRevLett.16.748}. The main open question about these particles is related to their acceleration mechanisms. Possibilities (below the GZK cutoff) are the Fermi mechanism \citep{1954ApJ...119....1F}, diffuse shock acceleration \citep{federico,ANCHORDOQUI20191}, among many others (for reviews, see, e.g., \cite{vietri, nagano, cronin, kotera2011, selvon, sokolsky2020, 10.1093/mnras/sty2936,2023MNRAS.522.4955H, 10.1093/mnras/stac3589}). 
Concerning sources of UHECRs, after  ICECUBE's ultra-high energy neutrinos, intensive work went in the direction of AGNs \citep{icecubea,icecubeb} (for a review, see \cite{universe8110607}). The main mechanisms for the production of UHECRs in AGNs are related to shock acceleration processes in the jets and in the magnetospheres of rotating BHs \cite{coimbra2018,coimbra2020,tursunov2020,coimbra2022}.

It is still debatable if LIGO BHs are primordial \citep{clesse,jedamzik,carr} or have stellar origins \citep{vattis,spera}. 
Despite being an open problem, the stellar origin of LIGO BHs is generally mostly accepted \citep{spera}. 
Here we assume that LIGO-Virgo BHs have a stellar origin and that they could inherit properties of the progenitor star, such as part of the original electromagnetic fields, or even that they could be charged. 
We also take as a working hypothesis that the superposition of BH electromagnetic fields during the merger brings immense energies to the configuration and UHECRs could be a byproduct of particles being accelerated at the BH stable orbits. If those charged BBHs have magnetospheres, then they could induce electromagnetic high-energy astrophysical phenomena. In this respect, for example, simulations of two BHs in a magnetically dominated plasma suggest that this system could generate an electromagnetic structure similar to those inferred in, e.g., collimated jets (for more details see \cite{milo,oneill,palenzuela,palenzuela2010,bode,moesta,giacomazzo,gold,ascoli}).

We investigate the spacetime energy extraction using the Blandford-Znajek (BZ) process \citep{blandford}, the Banados-Silk-West (BSW) effect \citep{banados}, and the magnetic Penrose process (MPP) \citep{tursunov2020} in the context of GW sources. The original BSW effect proposes that the collision of two neutral classical particles freely falling onto extremal Kerr BHs with mass $M$ [$a/M=1$, where $a\equiv J/M$ is the rotation parameter and $J$ is the BH's angular momentum] could produce ultra-high center-of-mass energies ($E_{\rm{c.m.}}$). The BSW effect was also evoked in models with static, charged, or rotational BHs with colliding neutral or charged test particles (see, e.g., \cite{jacobson,wei,harada1,kimura,zasla2011,frolov,igata,gosh,harada2,hussain,shiose,zhang2016,guo2016,lupsasca,coimbra2018,coimbra2020,coimbra2022,liberati2022}; for a more general review on the collisional Penrose process, see, e.g., \citep{jeremy2018}). Here we model GW BHs close to the full merger limit as either Kerr or Kerr-Newman. Colliding particles are classical charged ones at the BH stable orbits and, for simplicity, we assume that particle geodesics do not present backreaction effects. Further details about the models are provided in the Supplemental Material, hich includes Refs. \citep{teukolsky,2014PhRvD..90h4032L} about it.

In addition to BBHs endowed with a small charge, an BH-NS binary would be another physical scenario from which UHECRs could emerge. Indeed, the NS could provide the background magnetic field that could change the trajectories of charged particles around a BH and lead to the realization of the BSW effect. Given that BH-NS binary systems are a reality in nature (as clear from GW detections), the opportunity presents itself to also study some consequences of particle acceleration in this context.

\noi {\bf {\it GW source aspects.---}} The last update of the GWTCs (plus some reported last detections) points to more than 90 detected events, mostly being BBH mergers, but some also being BH-NS binaries or even BNSs \citep{spera}. Table I of the Supplemental Material summarizes the properties of most mergers (which includes Refs. \citep{ligo1,ligo2} about it), associated with the highest rotation parameter ($\chi \equiv a/M=J/M^2$) up to date. The largest $\chi$ of a BBH merger is around 0.9 and the majority of merged BHs have $\chi$ around 0.7-0.8. For events involving NSs, $\chi$ is smaller, up to around 0.4. The merging black hole masses are typically around several dozen of solar masses. But there are noticeable outliers, with final masses above 100-150 $M_{\odot}$. In our UHECR estimates, we will cover the mass and rotation parameters most common to the LIGO-Virgo-KAGRA (LVK) catalogs.

\noi {\bf {\it Results.---}} The aim of this letter is to demonstrate that ordinary magnetic fields on the surfaces of NSs ($\sim 10^{10}-10^{13}$ G) and very small charge-to-mass ratios of black holes could efficiently accelerate particles to ultra-high energies via the BSW mechanism. This results in a maximum particle energy of
$E_{max} \sim 10^{20}\left( \frac{f(a/M,B, \{L_i,{\cal E}_i\})}{10^5}\right)\left(\frac{M}{100 M_{\odot}}\right)\left(\frac{a/M}{0.8}\right)\left(\frac{B}{10^{11} G}\right)$ eV.  Here, $B$ represents the strength of the magnetic field on the BH, and $f(a/M, B, \{L_i,{\cal E}_i\})$ is the form factor of the center-of-mass energy of particles (each one with an angular momentum $L_{i}$ and energy ${\cal E}_i$; $\{ L_{i}, {\cal E}_{i}\}$ is the set of all of them) accelerated by the BSW mechanism, defined below. This mechanism accelerates particles at the BH’s Innermost Stable Circular Orbit (ISCO). 

In particular, the center-of-mass energy of a two-particle system with mass $m_0$, for a source with a given $g_{\mu\nu}$ metric, is \citep{banados}

\be
\frac{E_{c.m.}}{\sqrt{2} m_0} =\sqrt{1-g_{\mu \nu}u^\mu_{(1)}u^\nu_{(2)}} \equiv f(a/M,B,L_1,L_2,{\cal E}_1,{\cal E}_2),
\ee

\noi where $u^\mu_{(1)}$ and $u^\nu_{(2)}$ are the four-velocities ($u^\alpha = \dot{x}^\alpha \equiv dx^{\alpha}/d\tau$, with $\tau$ the proper time) of each particle. For charged particles with charge $q$ in an approximately constant magnetic field, $\dot{x}^{\mu}_{(i)}$ can be calculated from the integration of the equations of motion, namely,
\begin{equation}
    \ddot{x}^\mu + \Gamma^\mu_{\alpha \beta} \dot{x}^\alpha\dot{x}^\beta = \frac{q}{m_0}F^\mu_\nu \dot{x}^\nu,
\end{equation}
where $F_{\mu\nu} =\partial A_{\nu}/\partial x^{\mu} - \partial A_{\mu}/\partial x^{\nu}$, $\Gamma^\mu_{\alpha \beta}$ the Christoffel symbols associated with $g_{\mu\nu}$ and $A_{\mu}=(B g_{\phi\phi}/2)\delta^{\phi}_{\mu}$. The integration could be more easily done with the use of the normalization condition $g_{\mu\nu}\dot{x}^{\mu}\dot{x}^{\nu}=-1$ and the constants of the motion [$\mathcal{E}=-g_{t\mu}(m_0 \dot{x}^\mu + q A^\mu)$ and $L=-g_{\phi\mu}(m_0 \dot{x}^\mu + q A^\mu)$], associated with the symmetries of the Kerr spacetime. For further information, see the Supplemental Material. The c.m. energy drives particles to leave the binary system and modulates the maximum energy $E_{max}$ of particles of a given source.\footnote{The presence of a BH and its magnetic field can affect the maximum energy that protons can achieve. A helpful method for establishing the relationship between the magnetic field and the mass of the black hole was described in \cite{Ptitsyna_2010,PhysRevD.96.023008,tursunov2020}. This technique uses the Hillas condition to calculate the maximum energy based on a particular condition while taking into account limitations, interaction losses, and the shape of the source.} Only particles with a Larmor radius higher than the source radius are considered in this context.
\footnote{The Larmor radius is the characteristic length scale that defines the propagation of an ultra-high energy particle with energy ${\cal E}$ and charge $Ze$ in a magnetic field $B$. A larger Larmor radius corresponds to a trajectory with a smaller curvature. This constraint offers a qualitative criterion for identifying potential sources of UHECRs by examining the largest values of the product $BR$ \cite{ANCHORDOQUI20191}.}

When an NS is close enough to the BH, the magnetic field strength around the BH's event horizon should be a fraction of those at the NS's surface. Assuming that the NS's magnetic field is approximated by a dipolar field, which decreases $\propto r^{-3}$, one can clearly see that magnetic fields of $10^{7}-10^{12}$ G happen when the distance of the NS to the BH region where collisions take place is smaller than around ten times the radius of the star. However, such a distance from the center of the star is approximately the event horizon radius, $r_h$, of a BH with around $100M_{\odot}$ and $a/M\approx 0.8$ (or around two times $r_h$ of BHs with dozens of solar masses and the same rotational parameters). In addition, in the case of charged BHs with a charge-to-mass ratio $\alpha=10^{-5}$, in the Supplemental Material, we show that the local magnetic fields around twice the BH's event horizon are $\sim 10^{11}-10^{12}$~G (when electric fields are concerned, they are $\sim 10^{11}-10^{12}$ statVolt/cm for the same distance to the BH). Thus, the effective distances of the centers of mass of the BHs or the BH and the NS would be around (2-5)$r_h$ for the magnetic BSW effect to possibly become relevant (given the ISCOs involved), roughly meaning (by means of Kepler's Third Law with $\sim 10\%-20\%$ corrections due to PN effects \citep{2014LRR....17....2B}) that the characteristic GW frequencies would range between 100 and 400 Hz. Although the presence of electric fields could also efficiently accelerate charged particles, we will not focus on this case in this work, but rather on the effect a magnetic field could have on the change of geodesics and their center-of-mass energies. (The action of an electric field would just increase the energies of charged particles and make it easier for them to escape.)  In addition, charge-to-mass ratios like the ones we take into account are much smaller than those that would significantly change the GW waveforms assuming no charge at all \citep{2021PhRvL.126d1103B,2021arXiv210407594W,2022PhRvL.128g1101B}. 

\begin{figure}
\centering
\includegraphics[width=8 cm]{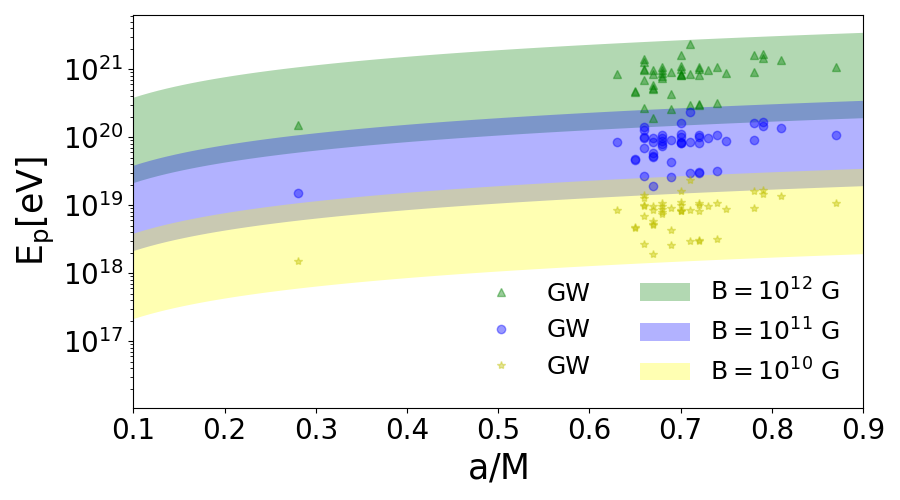}
\caption{Energy of a proton ($E_p=E_{max}$) as a function of the rotation parameter $a/M$ for values of the magnetic field of normal pulsars. The points correspond to the BH rotation parameters and masses present in the LIGO-Virgo-KAGRA catalogs (summarized in the Supplemental Material). Each marker/colored zone corresponds to a given value of $B$. We have chosen for all figures of this letter $L_1= 2.63245Mm_p$, $L_2= 2.08944Mm_p$, and ${\cal E}_1/m_p={\cal E}_2/m_p=1$. 
The borders of a colored zone (for a given $B$) are obtained from the largest and smallest BH masses. An overlap between two colored zones arises because of $E_p\propto M$ and the broad range of observed BH masses.
}
\label{fig1}
\end{figure}

\begin{figure}
\centering
\includegraphics[width=8 cm]{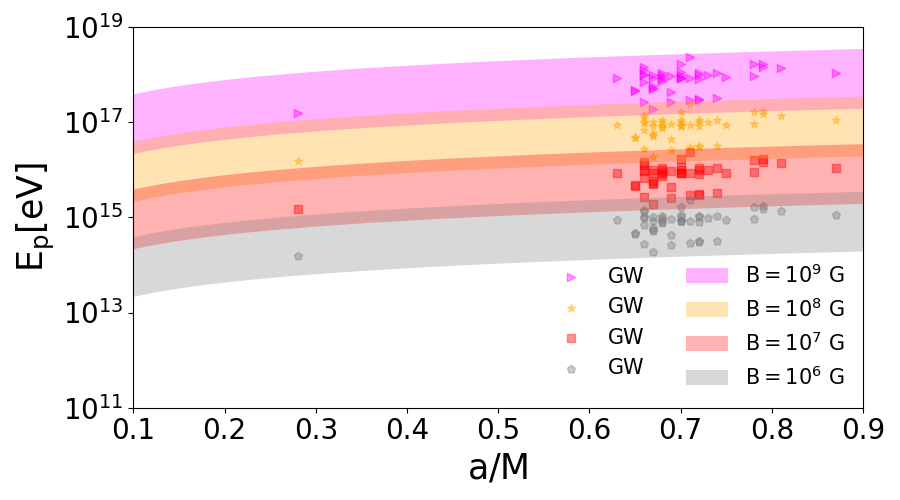}
\caption{Energy of a proton 
as a function of the rotation parameter for low values of the magnetic field. The points in the plot have the same meaning as in Fig. \ref{fig1}. }
\label{fig2}
\end{figure}

\begin{figure}
\centering
\includegraphics[width=8 cm]{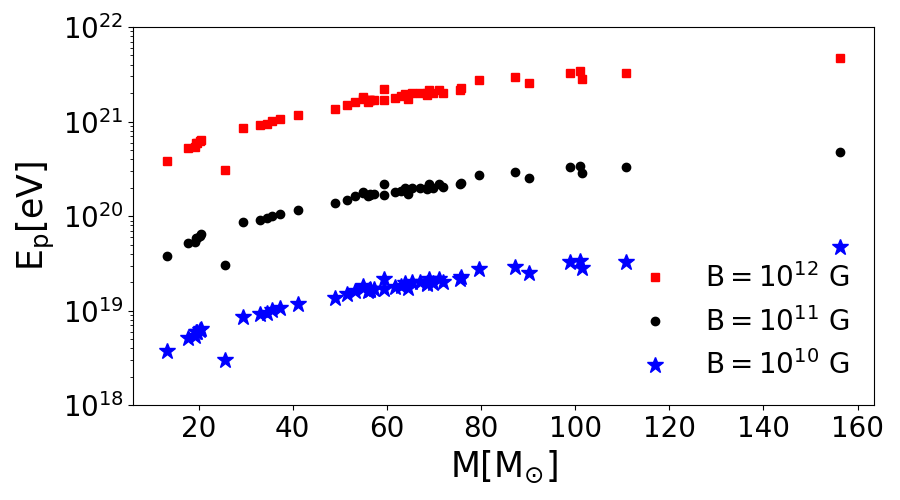}
\caption{Energy of a proton as a function of the GW mass and spin for different values of the magnetic field. For each value of $B$, the points correspond to the BH masses in the LIGO-Virgo-KAGRA catalogs.}
\label{fig3}
\end{figure}

\begin{figure}
\centering
\includegraphics[width=8 cm]{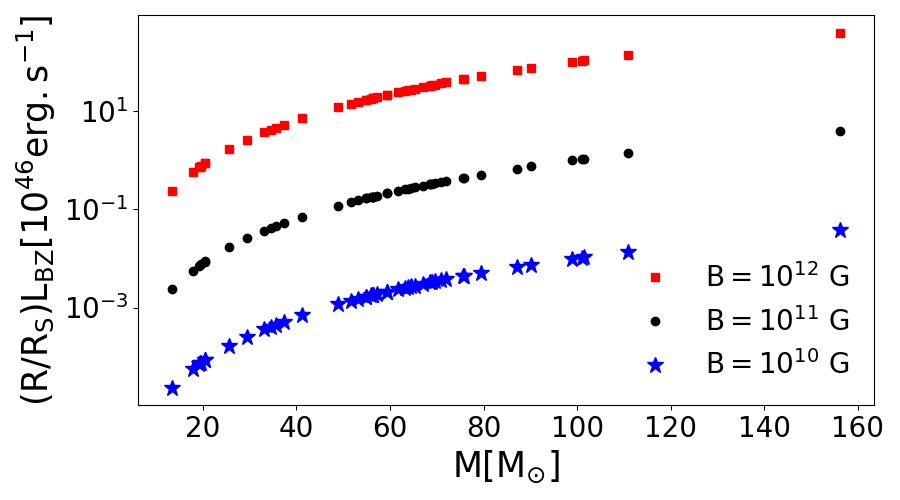}
\caption{Poynting flux as a function of the GW mass and spin for different values of the magnetic field $R$ is the orbital radius as the Schwarzschild radius $R_S$ \citep{kotera2016}. The points here have the same meaning as in Fig. \ref{fig3}.
}
\label{fig4}
\end{figure}

Figures \ref{fig1} and \ref{fig2} show the estimation of the energy of escaping protons $E_p\equiv E_{\rm{max}}$ (with mass $m_p$ and the result of the collision of two protons with angular momenta $L_1= 2.63245Mm_p$ and $L_2= 2.08944Mm_p$ and energies ${\cal E}_1/m_p={\cal E}_2/m_p=1$) \footnote{The conservative range of the dimensionless quantity $\ell \equiv L/(Mm_0)$ for the BSW effect is $-2(1+\sqrt{1+a/M})<\ell <2(1+\sqrt{1-a/M})$ \citep{banados}. For additional details, see the Supplemental Material.} for the GW parameter events as a function of the rotation parameter considering fields ranging from $10^{6}$ G to $10^{12}$ G (in multiples of 10)  and equatorial motions just for simplicity. Each marker or colored zone corresponds to a specific $B$ value. The borders of each colored zone are determined by the maximum and minimum BH masses from the LVK catalogs (approximated by $160M_{\odot}$ and $10M_{\odot}$, respectively; see Figs. \ref{fig3} and \ref{fig4}). An overlap between two colored zones with different $B$s occurs due to $E_p=E_{max}\propto M$ and the wide range of BH masses from GW events. As can be seen in both figures, magnetic fields approximately larger than $\sim 10^{10}$ G can lead to ultra-high energy particles. Figures \ref{fig3} and \ref{fig4} show the energy of a proton and the Poynting flux as a function of the BH mass (as appearing in the GW catalogs) for magnetic fields around $10^{10}-10^{12}$ G and same set $(L_1,L_2,{\cal E}_1,{\cal E}_2)$ as in Fig. \ref{fig1}. The Poynting flux can be preliminarily calculated using an estimate given by Lyutikov \cite{PhysRevD.83.124035, kotera2016} as $L_{BZ} \sim 10^{46} M B (R\slash R_{S}) \; \mathrm{erg.s^{-1}}$, where $M$ is the final mass of the black hole, $B$ is the strength of the external magnetic field, and the Schwarzschild radius $R_S$ is equal to the orbital radius R. 

It turns out that the proton energy and Poynting flux are almost insensitive to the rotation parameter ($a/M$) for the above magnetic fields and that suggests that binary coalescences should be able to accelerate particles to ultrahigh energies. 
From the above figures, one can also see that larger BH masses lead to higher proton energies, although the difference is just around an order of magnitude.

\noi {\bf {\it Discussion.---}} 
\noindent The BSW effect suggests that UHECRs could emerge from neutral BHs. However, the conditions for this are too demanding and are very unlikely to occur in nature \citep{2009PhRvL.103w9001B}. 
The presence of magnetic fields is a natural way to extend it as they affect the trajectories of charged particles and are ubiquitous in astrophysics. In the case of supermassive BHs, magnetic fields come from the accretion discs (plasma) surrounding them and are also present in the jets they produce. Radio galaxies, for example, are powerful sources presenting jet termination shock and large-scale lobes where particles can be accelerated up to high energies \cite{universe8110607, BELL201356, 2022ApJ...925...42D, 10.1093/mnrasl/sly099}. Magnetic fields associated with gravitational wave BHs, on the other hand, should be mostly related to their charges. We have shown that tiny charge-to-mass ratios of around $10^{-5}$ could already be enough to produce magnetic fields around $10^{10}-10^{12}$ G in the neighborhoods of event horizons of BHs, significantly affecting the trajectories of charged particles there. As a result, it is possible to significantly decrease the dependence of the original BSW effect on the rotation parameter ($a/M$) of the BH and largely increase its window of action. Indeed, we have found that for magnetic fields larger than $10^{10}$ G, common for NSs and present for tiny charge-to-mass ratios of BHs, the effect would appear for virtually any possible $a/M$.

In the case of charged BHs, a residual accretion disk in the binary is still necessary to provide the charged particles that could collide and allow the BSW effect. This issue of an accretion disk is still an open problem in the context of GWs. The reason is that the disk would only have a fraction of the total mass of the binary. Indeed, the mass transfer from the companion or surrounding gas onto the BH is often not very efficient, and a significant fraction of the material may not make it to the accretion disk. Additionally, the gravitational interaction between BHs in a binary system can disrupt or disturb the accretion process, limiting the growth of the disk. 

The non-detection of UHECRs from binary coalescences within the context of the BSW mechanism may suggest that residual disks are not present in BBH systems. Another possibility is that the density of charged particles in the disks is not large enough to evidence those pairs that collide near the event horizons of BHs and lead to high center-of-mass energy. A third possibility is that BHs are not charged. To gain a better understanding on these possibilities, further investigation is needed. It is necessary to study the presence and properties of accretion disks in BBH systems, their mass and composition, and their potential role in the BSW effect and GW production. 

A more feasible scenario for UHECRs would be BH-NS binaries. In this case, the magnetic field is naturally provided by the NS, and it can reach strengths of around $10^{7}-10^{12}$ G when the NS is at a distance ranging from $10R$ to almost touching the event horizon radii of the BH. However, this situation only occurs very close to the merger, and test particles could be supplied by the NS magnetosphere or even by the NS itself if it begins to disrupt due to tidal interactions with the BH. In this case, it seems possible to have a large enough number of particles to render the BSW effect relevant. For the case of BH-NS coalescences, the upper limit to the number of high-energy events--associated with near-extreme Kerr BHs ($a/M\rightarrow 1$)--can be estimated as follows. From magnetospheric aspects of an NS \cite{julian,mestel} whose surface is at $\delta r$ of the ISCO of a BH, the charge-to-current ratio ($t_r$) of particles colliding inside a column of radius $r_{ISCO} (\approx M)$ is $t_r\sim r_{ISCO}^2 \delta r/(v_{ISCO}R^2)$, where $R$ is the NS's radius and $v_{ISCO}$ the local speed of particles at the ISCO. For the typical values $\delta r =(10^2-10^5)$ cm, $R=12$ km, $M=10M_{\odot}$, and $v_{ISCO}=0.3c$, we have that the number of events per second would be $\sim 1/t_r= 10^4-10^7$. For BHs with the above masses and GW frequencies in the range 100-400 Hz, the merger time can be estimated as $t_{merger}=(0.01-2)$s \citep{2018PhRvD..97b3016A}.\footnote{Once created, high-energy particles would almost freely leave the BH-NS binary because this would occur before the merger. These particles, moving at speeds close to the speed of light, could travel distances around $10^2-10^4$ NS radii within $t_{merger}$.} Thus, the total number of high-energy events before the merger is $\sim t_{merger}/t_r=10^2-10^7$. For BBH mergers with accretion disks, one should use the cross-section $\sigma$ of charged particles \cite{goldston1995,hazeltine2018} crossing the ISCO. For typical plasma parameters in the context of BHs with electron number density $n_e\sim 10^{14}$cm$^{-3}$ and BHs with $10M_{\odot}$, the cross-section of heavy particles \cite{goldston1995,hazeltine2018} can be estimated as $\sigma \sim  10^{-21}$cm$^2$. For the speeds of particles crossing the ISCO ($v_{ISCO}\approx 0.3 c$), the upper limit to the number of high-energy events per second would be $\sim n_ev_{ISCO}\sigma=10^{2}$. Thus, the maximum number of events before the merger would be $\sim 10^{-2}-10^2$. However, accretion disks of low mass BHs could have larger values of $n_e$, e.g., $10^{17}-10^{20}$ cm$^{-3}$ \citep{2016MNRAS.462..751G}. In this case, the number of high-energy events with respect to a conservative $n_e$ would be increased by a factor of $10^3-10^6$, meaning that up to millions of them could happen before the merger of BBHs. For further details, see the Supplemental Material.

The results of Figs. \ref{fig1} and \ref{fig2} concern proton energies at the source. However, UHECRs are related to particles detected on Earth, and thus they need to escape and propagate from the source to the observer. Regarding the escape, differently from the case of near-extremal Kerr BHs \citep{2012PhRvL.109l1101B}, the presence of the magnetic fields facilitates that (the larger $B$ is, the more probable that becomes). \footnote{In the presence of a magnetic field, particles are expelled at positions slightly shifted from the BH ISCO. Fields could be large there depending on the distance of the BH's companion to such positions.}. 
When it comes to cosmic ray propagation, it would imply energy losses. 
Several energy-loss mechanisms have to be taken into account as pair production and photodisintegration. Protons and nuclei may undergo pair production when the cosmic microwave background (CMB) photons are involved. In the case of UHE nuclei, they can interact with infrared, optical, and ultraviolet photons. In addition to all the previously mentioned factors, the binary's characteristics, the specific details of the merger process, and the acceleration mechanisms at the source can all have a substantial impact on the energy spectrum of UHECRs observed on Earth. The photo-decay of nuclei and the process of pair formation lead to a steepening of the observed spectrum: the cut-off in the observed spectrum above some $10^{19}$ eV. In the photodesintegration process, a nucleus of atomic mass number $A$ loses one or more nucleons due to the interaction with background radiation and the most important process is the emission of one nucleon \citep{ANCHORDOQUI20191}. Consequently, we lose relevant particle information owing to energy losses and deflections by magnetic fields. However, in the coming years with AugerPrime, we expect the detection of more energetic particles with greater accuracies in mass, allowing a study of sources and acceleration processes. The AugerPrime upgrade will increase the sensitivity to primary cosmic ray composition above $\sim 10^{19}$ eV. It is now the detector system with the highest exposure for UHECR \citep{ANASTASI2022167497}.

As a binary system with a BH nears merger, deviations from the Kerr metric become relevant, requiring numerical relativity. These deviations affect ``$f$'' and thus the quantities in Figs. \ref{fig1}, \ref{fig3}, and \ref{fig4}. For first estimates, one can take a BBH spacetime and it can be modeled as a superposition of two Kerr spacetimes with boost corrections for high BH speeds \cite{combi2021}. In the absence of a magnetic field, the gravitational potential of a binary is larger than that of a single Kerr BH, resulting in decreased c.m energies. This decrease is roughly proportional to the relative change in the $g_{tt}$ metric component. For two BHs of the same mass separated by $20M\equiv 10r_h$, the relative change in $g_{tt}$ at the c.m. location (at $5r_h$ from a BH's center) is $|1-g_{tt}^{\rm{BBH}}(5r_h)/g_{tt}^{\rm{Kerr}}(5r_h)|\sim 10\%$ \cite{combi2021}. In the presence of magnetic field, the c.m. energy and related observables decrease similarly. However, that happens in a nonlinear manner due to the complex interplay among the magnetic field, metric and particle parameters, and precise calculations are needed. The effects on c.m. energies of evolving event horizons and magnetic field dynamics in time-dependent binary system spacetimes also require careful study.

\noi {\bf {\it Acknowledgements.---}} We thank the anonymous referees for the relevant suggestions which helped improve this work. The research of R.C.A. is supported by Conselho Nacional de Desenvolvimento Cient\'{i}fico eTecnol\'{o}gico (CNPq) (310448/2021-2), FAPESP (2021/01089-1) and Fundação Araucária (698/2022). She also thanks for the support of L'Oreal Brazil, with the partnership of ABC and UNESCO in Brazil.  J.G.C. is grateful for the support of FAPES (1020/2022, 1081/2022, 976/2022, 332/2023), CNPq (311758/2021-5), and FAPESP (2021/01089-1). C.H.C.A., J.G.C., and R.C.A. gratefully acknowledge the financial support of the ``Fen\^omenos Extremos do Universo" of the Funda\c{c}\~ao Arauc\'aria. J.P.P. is thankful for the support of FAPES through grant No. 04/2022. We acknowledge the National Laboratory for Scientific Computing (LNCC/MCTI, Brazil) for providing HPC resources of the SDumont supercomputer, which have contributed to the research results reported within this paper. URL: https://sdumont.lncc.br.

\appendix

\section{Kerr BHs}\label{sec:kerr}

A Kerr BH is described by two parameters: a mass $M$ and an angular momentum $J$ (here represented by $a=J/M$, that is, the angular momentum per unit mass). The Kerr line element describes a stationary spacetime with axial symmetry and, in Boyer-Lindquist coordinates, it is written as \citep{teukolsky}

\be\label{eq:kerr}
ds^2 = g_{tt}dt^2 + 2g_{t \phi}dtd\phi + g_{rr}dr^2 + g_{\theta \theta}d\theta^2 + g_{\phi \phi}d\phi^2,
\ee

\noi with

\be
g_{t t} = -\left(1-\frac{2M r}{\Sigma} \right),
\ee

\be
g_{t \phi} = -\frac{2aMr\sin^2\theta}{\Sigma},
\ee

\be
g_{r r} = \frac{\Sigma}{\Delta},
\ee

\be
g_{\theta \theta} = \Sigma,
\ee

\be
g_{\phi \phi} = \frac{(r^2+a^2)^2-a^2\Delta \sin^2\theta}{\Sigma}\sin^2\theta,
\ee

\noi and where $\Sigma=r^2+a^2\cos^2\theta$ and $\Delta = r^2+a^2-2Mr$. The event horizon is located at $r_H = M + \sqrt{M^2 - a^2}$. 
The ergoregion of a Kerr-BH is described by $r_H < r < r_E(\theta)=M+\sqrt{M^2-a^2\cos^2\theta}$.

When considering the motion of particles only under the action of gravity, the conserved quantities along geodesics play a fundamental role. For the Kerr spacetime, the energy and angular momentum relative to the axis of symmetry of a particle are conserved quantities. 
Considering the motion of neutral or charged particles near rotating BHs in a background described by (\ref{eq:kerr}),
the conserved quantities are attached to Killing vectors $\xi_{(t)}=\xi_{(t)}^\mu\partial_\mu=\frac{\partial}{\partial t}$ and $\xi_{(\phi)}=\xi_{(\phi)}^\mu\partial_\mu=\frac{\partial}{\partial \phi}$. The first one is related to the free-test particle energy conservation

\be\label{eq:killing1}
\mathcal{E}=-g_{t\mu}p^\mu, 
\ee

\noi and the other to the free-test particle angular momentum conservation

\be\label{eq:killing2}
L=-g_{\phi\mu}p^{\mu}.
\ee
For convenience, we define the dimensionless quantity $\ell \equiv L/(Mm_0)$, where $m_0$ is the test particle's mass. 

\section{Local electromagnetic fields of a Kerr-Newman BH}

In order to select the physically interesting cases for particle acceleration when BHs are charged, one needs to calculate the local values of electromagnetic fields of a general Kerr-Newman spacetime. It is the generalization of the metric given by Eq. \eqref{eq:kerr} due to the presence of a charge $Q$, and the metric components in the Boyer-Lindquist coordinate system read
\be
g_{tt}=-\left[1-\frac{2Mr-Q^2}{\Sigma}\right]\label{gttKN},
\ee
\be
g_{rr}=\frac{\Sigma}{\tilde{\Delta}},\label{grrKN}
\ee
\be
g_{\theta\theta}=\Sigma \label{g22KN},
\ee
\be
g_{\phi\phi}=\frac{ (r^2+a^2)^2- \tilde{\Delta}a^2\sin^2\theta }{\Sigma}\sin^2\theta\label{g33KN},
\ee
\be
g_{t\phi}= \frac{(Q^2-2Mr)a}{\Sigma}\sin^2\theta \label{gt3KN},
\ee
where $\tilde{\Delta}\equiv \Delta + Q^2$.

The local electromagnetic fields can be derived with the use of (orthonormal) tetrads ($e^a_{\mu}$). They are defined such that
\be
g_{\mu\nu}=e^{a}_{\mu}e^b_{\nu}\eta_{ab},\;\; e^a_{\mu}e^{\mu}_b=\delta^a_b,\;\; e^a_{\nu}e^{\mu}_a=\delta^{\mu}_{\nu}\label{tetrad}
\ee
with $\eta_{ab}\equiv \mbox{diag}(-1,1,1,1)$. Furthermore, the index $\mu$ (a) could be raised or lowered with $g_{\mu\nu}$ ($\eta_{ab}$). Local fields are related to projections onto a given tetrad. They can be extracted from the electromagnetic tensor $F_{\mu\nu}$ as $F_{ab}\equiv e^{\mu}_ae^{\nu}_bF_{\mu\nu}$, where
\be
F_{\mu\nu}=A_{\nu;\mu}-A_{\mu;\nu}=A_{\nu,\mu}-A_{\mu,\nu}\label{Fmunu},
\ee
with $A_{\mu}$ is the electromagnetic four-potential and the semicolon (comma) denotes the covariant (partial) derivative. For the Kerr-Newnam spacetime \citep{2014PhRvD..90h4032L}, 
\be
A_{\mu} = \left(\frac{Qr}{\Sigma},0,0, -\frac{Qr}{\Sigma}a\sin^2\theta\right)\label{Amu}.
\ee

For the metric given by Eqs. \eqref{gttKN}--\eqref{gt3KN}, it follows that
\be
e_{\mu}^0=\left(B_1,0,0,B_2 \right)=\left(\sqrt{-g_{tt}},0,0, -\frac{g_{t\phi}}{\sqrt{-g_{tt}}}\right)
\ee

\be
e_{\mu}^1=\left(0,B_3,0,0 \right)=\left(0,\sqrt{g_{rr}},0,0 \right)
\ee

\be
e_{\mu}^2=\left(0,0,B_4,0\right)=\left(0,0,\sqrt{g_{\theta\theta}},0 \right)
\ee

\be
e_{\mu}^3=\left(0,0,0,B_5\right)=\left(0,0,0,\sqrt{ g_{\phi\phi} +\frac{(g_{t\phi})^2}{(-g_{tt})}}\right)
\ee

The inverse of these tetrad components are
\be
e^{\mu}_0=\left(A_1,0,0,0 \right)=\left(\frac{1}{B_1},0,0,0 \right)
\ee

\be
e^{\mu}_1=\left(0,A_2,0,0 \right)=\left(0,\frac{1}{B_3},0,0 \right)
\ee

\be
e^{\mu}_2=\left(0,0,A_3,0\right)=\left(0,0,\frac{1}{B_4},0\right)
\ee

\be
e^{\mu}_3=\left(A_4,0,0,A_5\right)=\left(-\frac{B_2}{B_1B_5},0,0,\frac{1}{B_5}\right)
\ee

The non-null components of the electromagnetic fields in Boyer-Lindquist coordinates are
\be
E_r=\frac{Q(r^2-a^2\cos^2\theta)}{\Sigma^2}
\ee

\be
E_{\theta}=-\frac{2Qra^2\sin\theta\cos\theta}{\Sigma^2}
\ee

\be
B_{r}=\frac{2Qra(r^2+a^2)\sin\theta\cos\theta}{\Sigma^2}
\ee

\be
B_{\theta}=\frac{Qa(r^2-a^2\cos^2\theta)\sin^2\theta}{\Sigma^2}
\ee
Finally, the local fields (projected onto the orthonormal frame defined as above) are
\be
E_{\hat{r}}=A_1A_2E_r
\ee

\be
E_{\hat{\theta}}=A_1A_3E_{\theta}
\ee

\be
B_{\hat{r}}=A_3A_5B_r-A_3A_4E_{\theta}
\ee

\be
B_{\hat{\theta}}=A_2A_5B_{\theta}+A_2A_4E_{r}
\ee

If one takes the flat spacetime limit to the above equations ($r\rightarrow\infty$), one gets (up to third order in $r$)
\be
E_{\hat{r}}=\frac{Q}{r^2}
\ee

\be
B_{\hat{r}}=\frac{2Qa}{r^3}\cos\theta
\ee

\be
B_{\hat{\theta}}=\frac{Qa}{r^3}\sin\theta
\ee
which are the expected field components for a magnetic dipole moment $Qa$ and the electric field of point-like particle with charge $Q$.

\begin{figure*}
  \centering
   {\includegraphics[angle=0,width=0.4\textwidth]{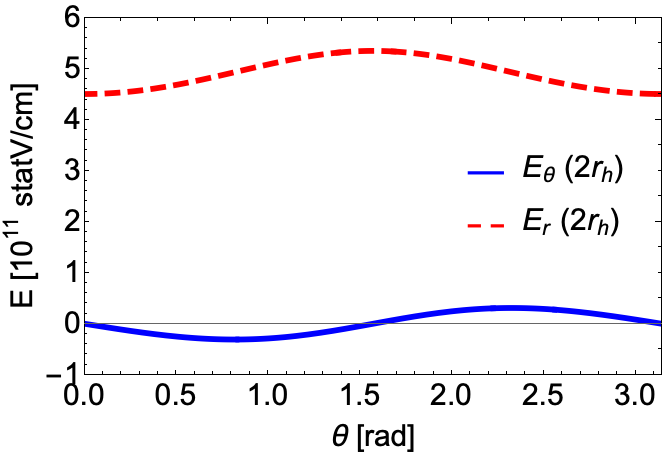}}
    {\includegraphics[angle=0,width=0.4\textwidth]{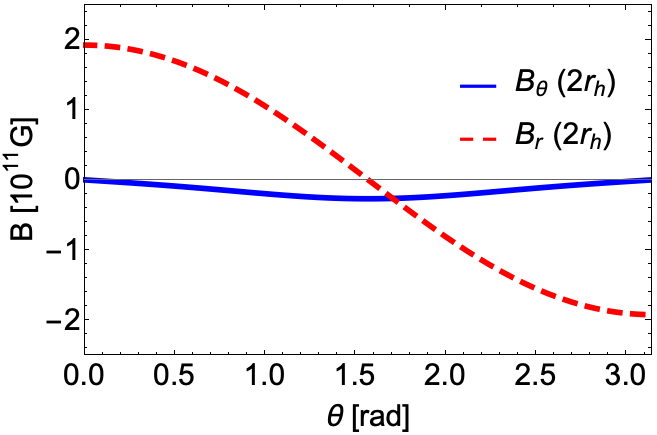}}
   \caption{Local electromagnetic field components as a function of the polar angle for $r=2r_h$ (twice the event horizon radius), $\alpha=10^{-5}$, $M\simeq 35 M_{\odot}$ and $a/M=0.7$.}
    \label{EBfields}
\end{figure*}

If one takes $Q/M\equiv \alpha = 10^{-5}$, one can find that near the event horizon of charged BHs, electromagnetic field components are smaller than $10^{11}-10^{12}$ G (statVolt/cm). This could be seen in Fig. \ref{EBfields} for different polar angles at the distance $r=2r_h$ in the case $M\simeq 35M_{\odot}$ and $a/M=0.7$. 

\section{The BSW effect in the presence of magnetic fields}

When particles transit the ergosphere of a Kerr or a Kerr-Newman black hole, they could have negative energy as determined by an observer at infinity \cite{press}. If the horizon captures these particles, it is possible to extract mass and angular momentum from the black hole. When several particles collide next to the horizon of an extremal black hole ($a/M \rightarrow 1$), they can reach arbitrarily high center-of-mass energy. This approach can be summarized as a particular Penrose collisional process known as the Ba\~nados-Silk-West (BSW) effect \cite{banados}.

Writing down the effective potential for particles following geodesics around a Kerr BH, one can choose values of $\mathcal{E}$ and $\ell$ in which the particles may be able to escape to infinity \cite{schn}, where the range of $\ell$ for so is $-2(1+\sqrt{1+a/M})<\ell <2(1+\sqrt{1-a/M})$ \citep{banados}. The center-of-mass energy of the two-particle system (each one with mass $m_0$) is given by \cite{banados}

\be\label{eq:ecm}
E_{c.m.} = \sqrt{2} m_0 \sqrt{1-g_{\mu \nu}u^\mu_{(1)}u^\nu_{(2)}},
\ee

\noi where $u^\mu_{(1)}$ and $u^\nu_{(2)}$ are the four-velocities ($u^\alpha = \dot{x}^\alpha \equiv dx^{\alpha}/d\tau$, where $\tau$ is the proper time) of each particle, properly normalized by $g_{\mu\nu}u^\mu u^\nu=-1$. This constraint, plus the symmetries connected to Eqs. \eqref{eq:killing1} and \eqref{eq:killing2}, allow the calculation of a complete set of geodesics to solve Eq. (\ref{eq:ecm}). It is expected that the main conditions for accelerating particles concern the BH spin since it influences the position of the innermost stable circular orbit (ISCO), which is related to the effective potential of the system. For example, consider the case of two particles, one of them at the ISCO and the other one coming from infinity, both with $\mathcal{E}_1/m_0 =  \mathcal{E}_2/m_0 =1$ and different angular momenta $\ell_1$ and $\ell_2$. For an extremal Kerr BH, the form of the center-of-mass energy at the ISCO reads

\be\label{eq:ecm_isco}
E_{c.m.}(r \rightarrow r_{ISCO}) = \sqrt{2} m_0 \sqrt{\frac{\ell_2 - 2}{\ell_1 - 2} + \frac{\ell_1 - 2}{\ell_2 - 2}}.
\ee

If one of the particles has the critical angular momentum $\ell = 2$, the center-of-mass energy blows up. In this case, the radial equation of motion (in geometric units) for each particle is $\dot{r}_{(1)} = (-2\mathcal{E}_1/m_0 + \ell_1)$ and $\dot{r}_{(2)} = (2\mathcal{E}_2/m_0 - \ell_2)$ \cite{banados}. 
Assuming the extremal values $\mathcal{E}_1/m_0=1$ and $\ell_1 = 2$ for the first particle, it indeed follows that $\dot{r}_{(1)}=0$ at the ISCO, i.e., the first particle is orbiting it. Choosing another critical angular momentum for the second particle, for instance, $\ell_2 = -2(1+\sqrt{2})$, we have that $\dot{r}_{(2)} = 2(\mathcal{E}_2/m_0+1+\sqrt{2})$, i.e., the particle approaches the ISCO with a non-zero radial velocity. The explicit value of the local speed of the second particle at the ISCO, in terms of $c$, is $v_{ISCO} \sim \frac{\mathcal{E}_2/m_0}{1+\sqrt{1+\mathcal{E}_2^2/m_0^2}}c$ \cite{press}. When $\mathcal{E}_2/m_0 = 1/\sqrt{3}$--another critical value for $\mathcal{E}/m_0$--it follows that $v_{ISCO} \sim 0.3 c$.

When magnetic fields are present, the geodesics change, as well as the position of the minimum effective potential. That, in turn, changes the positions where BSW may occur, or even alter the minimum values of $a/M$ that successfully trigger it. When a charged particle is orbiting the ISCO of a magnetized Kerr or Kerr-Newman BH, an incoming particle can collide with it, leading to three possible outcomes: (i) the charged particle is expelled to infinity; (ii) it could be trapped by the magnetized BH; (iii) it continues orbiting in the ISCO. The interaction of electromagnetic fields with charged particles may accelerate them (in the case of electric fields) or deflect them (in the case of magnetic fields). The presence of magnetic fields could also change the frame-dragging aspects of the spacetime near the ISCO of a BH. This is actually relevant since the cross-section for the capture of incoming particles can change--becoming larger--and that strongly depends on the black hole spin, the angular momentum of the particles \cite{chandrasekhar}, and the presence of magnetic fields themselves.

For a particle immersed in an approximately uniform magnetic field $B$ in a curved space, the equations of motion for it read 

\be\label{eq:lagrange}
\ddot{x}^\mu + \Gamma^\mu_{\alpha \beta} \dot{x}^\alpha\dot{x}^\beta = \frac{q}{m_0}F^\mu_\nu \dot{x}^\nu, 
\ee

\noi where 
$F_{\mu\nu}$ is given by Eq. \eqref{Fmunu}, $q$ the particle charge and $m_0$ the test particle mass.  
Using the Lorentz gauge $A^\mu_{\;;\mu}=0$, the only nonzero component of the four-potential is $A_\phi = B g_{\phi\phi}/2$. The four-momentum of the particle, in this case, is $p_\mu = m_0 u_\mu + q A_\mu$ and the conserved quantities in Eqs. (\ref{eq:killing1}) and (\ref{eq:killing2}) are written as

\be\label{eq:newkilling1}
\mathcal{E}=-g_{t\mu}(m_0 u^\mu + q A^\mu), 
\ee

\be\label{eq:newkilling2}
L=-g_{\phi\mu}(m_0 u^\mu + q A^\mu).
\ee

In the absence of backreactions, which we assume here, we can easily obtain the geodesic equations for charged particles. For simplicity, we focus on equatorial motions (i.e., $\dot{\theta}_{(i)}=0$). From Eqs. (\ref{eq:lagrange}), considering the conserved quantities in (\ref{eq:newkilling1}) and (\ref{eq:newkilling2}) ($\mathcal{E}_{(i)}/m_0=1$) and the normalization condition $u^\mu u_\mu = -1$, we have the following system of equations:

\begin{equation}
\dot{\phi}_{(i)} = \frac{g_{tt}L_i/m_0 + g_{\phi t} \mathcal{E}_i/m_0 -g_{tt}g_{\phi \phi} {\cal B}}{g_{tt}g_{\phi \phi} - (g_{t\phi})^2},
\end{equation}

\begin{equation}
\dot{t}_{(i)} = \frac{-\mathcal{E}_i/m_0-g_{t\phi}\dot{\phi}}{g_{tt}},
\end{equation}

\begin{equation}
\dot{r}_{(i)} = \sqrt{\frac{-1-g_{tt}(\dot{t}_{(i)})^2-2g_{t\phi}\dot{t}_{(i)}\dot{\phi}_{(i)}-g_{\phi\phi}(\dot{\phi}_{(i)})^2}{g_{rr}}},
\end{equation}

\begin{figure*}
  \centering
   {\includegraphics[angle=0,width=0.4\textwidth]{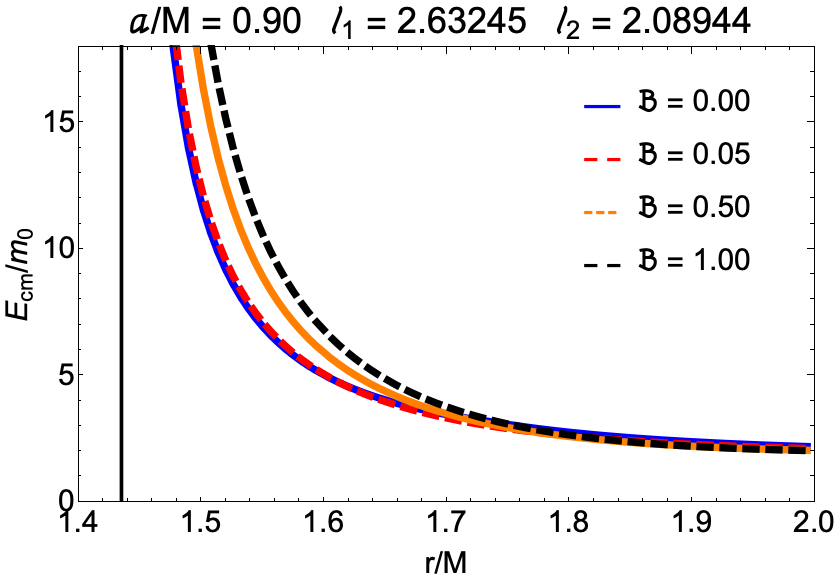}}
    {\includegraphics[angle=0,width=0.4\textwidth]{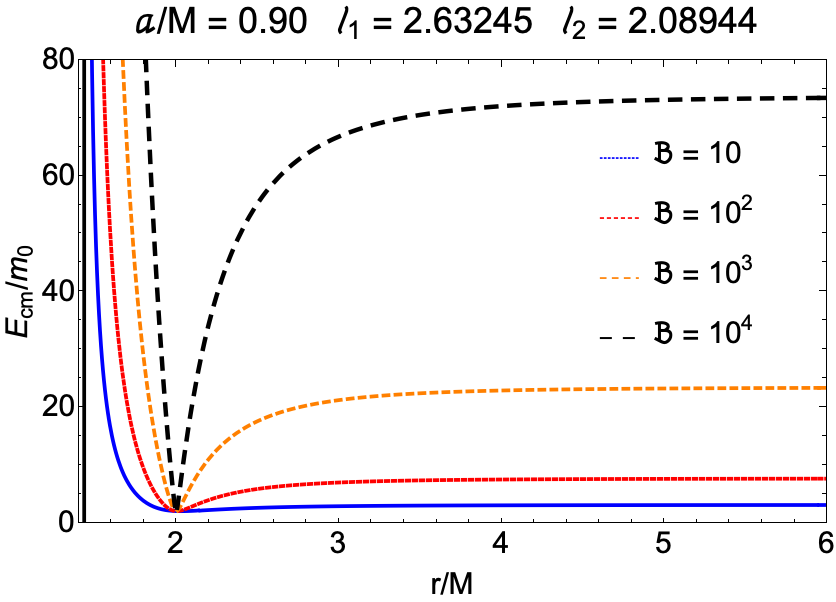}}\\
  {\includegraphics[angle=0,width=0.4\textwidth]{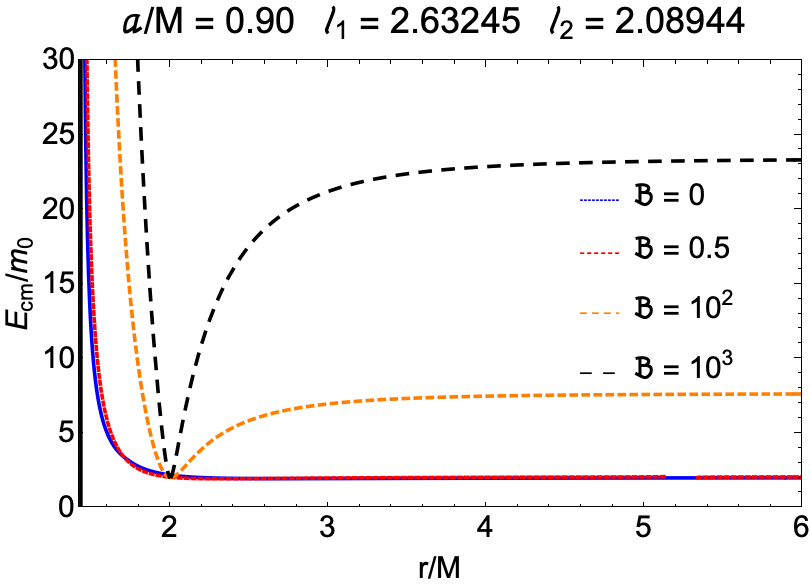}}
   \caption{$\mathrm{E_{cm}}$ values for $a/M = 0.9$ and different normalized magnetic field strengths (in C.T/m units).}
    \label{Ecm_09}
\end{figure*}

\begin{figure*}
  \centering
   {\includegraphics[angle=0,width=0.4\textwidth]{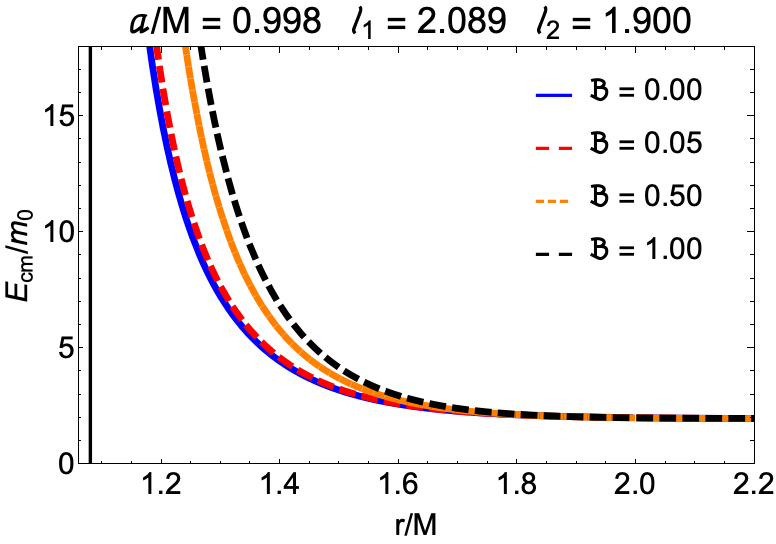}}
    {\includegraphics[angle=0,width=0.4\textwidth]{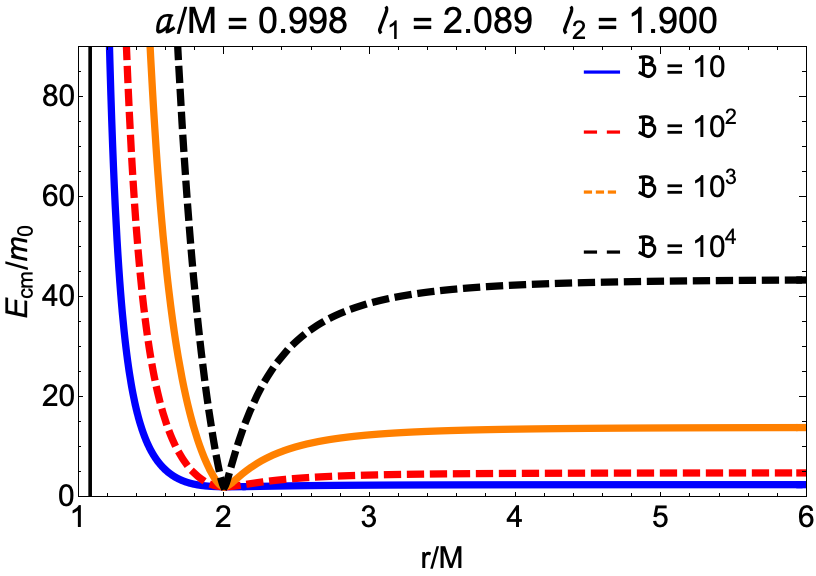}}
   \caption{$\mathrm{E_{cm}}$ values now for $a/M = 0.998$ and different normalized magnetic field strengths (in C.T/m units).}
    \label{Ecm_0998}
\end{figure*}

\noi  where ${\cal B} \equiv \frac{q B}{2m_0}$ is the normalized magnetic field strength. The center-of-mass collision energy of two particles is calculated from Eq. (\ref{eq:ecm}). Figs. \ref{Ecm_09} and \ref{Ecm_0998} show results of the $E_{c.m}$ for a set of possible particle angular momenta and black hole spins of near-extremal ($a/M=0.998$) BHs in the Thorne limit and also for BH spins of order $a/M \sim 0.9$. This last case is more akin to what is seen in GW catalogs containing the merger of black hole binaries. Particles are expelled to infinity at positions slightly shifted from the BH ISCO (black vertical line), and that depends on the intensity of $B$ (which also depends on the distance of the BH companion to such positions). The more intense $B$ is, the greater the chances of particles being expelled. This makes it easier for the effective occurrence of the BSW effect since particles could escape more easily from the system.

\section{LIGO-Virgo-KAGRA GW events}
Table \ref{table1} summarizes the redshifts, masses, and rotational parameters of GW events detected so far by the LVK collaboration, for $\chi_f\equiv a/M > 0.7$ and all the BH-NS events.

\begin{table*}
\centering
\renewcommand{\arraystretch}{1}
\caption{List of LIGO-Virgo-KAGRA sources with rotational parameter $\chi_f\equiv a/M\equiv J/M^2 > 0.7$ ($J$ is the BH's angular momentum) for the BH produced after the merger of BBHs/BH-NS/BNS \citep{ligo1,abbott2021a,abbott2021b,ligo2}. The list is ordered following the chronological detections. The masses of the BHs/NSs before the merger are denoted by $m_1$ and $m_2$, and $M_f$ is the mass of the final BH/NS. All masses are in solar mass (M$_\odot$) units. The redshift of a source is $z$. The events are classified here as binary of black holes (BBHs) or binary of black hole-neutron star (BH-NS) or binary of neutron stars (BNSs).} \label{table1}
    \begin{tabular}{llllllll}
    \hline
    \hline
Catalog & Event       & $z$ & $m_1$ & $m_2$  & $M_f$      & $\chi_f$  & Type       \\ \hline
GWTC-1 & GW151226    & 0.09$^{+0.04}_{-0.03}$  & 13.7$^{+8.8}_{-3.2}$& 7.7$^{+2.2}_{-2.5}$ &20.5$^{+6.4}_{-1.5}$ & 0.74$^{+0.07}_{-0.05}$ & BBH \\
GWTC-1 & GW170729    & 0.49$^{+0.19}_{-0.21}$& 50.2$^{+16.2}_{-10.2}$ &34.0$^{+9.1}_{-10.1}$ &79.5$^{+14.7}_{-10.2}$& 0.81$^{+0.07}_{-0.13}$ & BBH\\ 
GWTC-1 & GW170814 & 0.12$^{+0.03}_{-0.04}$ & 30.6$^{+5.6}_{-3.0}$ & 25.2$^{+2.8}_{-4.0}$ & 53.2$^{+3.2}_{-2.4}$ & 0.72$^{+0.07}_{-0.05}$ & BBH\\
GWTC-1 & GW170817$^*$  & 0.01$^{+0.00}_{-0.00}$ & 1.46$^{+0.12}_{-0.10}$ & 1.27$^{+0.09}_{-0.09}$&$\leq 2.8$& $\leq 0.89$ & BNS\\
GWTC-1 & GW170823    & 0.35$^{+0.15}_{-0.15}$& 39.5$^{+11.2}_{-6.7}$ &29.0$^{+6.7}_{-7.8}$ &65.4$^{+10.1}_{-7.4}$& 0.72$^{+0.09}_{-0.12}$ & BBH\\ 
GWTC-2 & GW190424$\underline{\;\;}$180648 & 0.39$^{+0.23}_{-0.19}$ & 40.5$^{+11.3}_{-7.3}$ & 31.8$^{+7.6}_{-7.7}$ & 68.9$^{+12.4}_{-10.1}$ & 0.74$^{+0.09}_{-0.09}$ & BBH\\
GWTC-2 & GW190517$\underline{\;\;}$055101 & 0.34$^{+0.24}_{-0.14}$ & 37.4$^{+11.7}_{-7.6}$ & 25.3$^{+7.3}_{-7.0}$&59.3$^{+9.1}_{-8.9}$& $0.87^{+0.05}_{-0.07}$ & BBH\\
GWTC-2 & GW190519$\underline{\;\;}$153544   & 0.44$^{+0.25}_{-0.14}$  & 66.0$^{+10.7}_{-12.0}$& 40.5$^{+11.0}_{-11.1}$ &101.0$^{+12.4}_{-13.8}$ & 0.79$^{+0.07}_{-0.13}$ & BBH \\
GWTC-2 & GW190521  & 0.64$^{+0.28}_{-0.28}$& 95.3$^{+28.7}_{-18.9}$ &69.0$^{+22.7}_{-23.1}$ &156.3$^{+36.8}_{-22.4}$& 0.71$^{+0.12}_{-0.16}$ & BBH\\ 
GWTC-2 & GW190521$\underline{\;\;}$074359 & 0.24$^{+0.07}_{-0.10}$ & 42.2$^{+5.9}_{-4.2}$ & 32.8$^{+5.4}_{-6.4}$ & 71.0$^{+6.5}_{-4.4}$ & 0.72$^{+0.05}_{-0.07}$ & BBH\\
GWTC-2 & GW190527$\underline{\;\;}$092055    & 0.44$^{+0.34}_{-0.20}$ & 36.5$^{+16.4}_{-9.0}$ & 22.6$^{+10.5}_{-8.1}$&  56.4$^{20.2}_{-9.3}$ & 0.71$^{+0.12}_{-0.16}$ & BBH\\
GWTC-2 & GW190620$\underline{\;\;}$030421    & 0.49$^{+0.23}_{-0.20}$& 57.1$^{+16.0}_{-12.7}$ &35.5$^{+12.2}_{-12.3}$ &87.2$^{+16.8}_{-12.1}$& 0.79$^{+0.08}_{-0.15}$ & BBH\\ 
GWTC-2 & GW190706$\underline{\;\;}$222641 & 0.71$^{+0.32}_{-0.27}$ & 67.0$^{+14.6}_{-16.2}$ & 38.2$^{+14.6}_{-13.3}$ & 99.0$^{+18.3}_{-13.5}$ & 0.78$^{+0.09}_{-0.18}$ & BBH\\
GWTC-2 & GW190719$\underline{\;\;}$215514 & 0.64$^{+0.33}_{-0.29}$ & 36.5$^{+18.0}_{-10.3}$ & 20.8$^{+9.0}_{-7.2}$&54.9$^{+17.3}_{-10.2}$& $0.78^{+0.11}_{-0.17}$ & BBH\\
GWTC-2 & GW190720$\underline{\;\;}$000836    & 0.16$^{+0.12}_{-0.06}$& 13.4$^{+6.7}_{-3.0}$ &7.8$^{+2.3}_{-2.2}$ &20.4$^{+4.5}_{-2.2}$& 0.72$^{+0.06}_{-0.05}$ & BBH\\ 
GWTC-2 & GW190727$\underline{\;\;}$060333 & 0.55$^{+0.21}_{-0.22}$ & 38.0$^{+9.5}_{-6.2}$ & 29.4$^{+7.1}_{-8.4}$ & 63.8$^{+10.9}_{-7.5}$ & 0.73$^{+0.10}_{-0.10}$ & BBH\\
GWTC-2 & GW190728$\underline{\;\;}$064510 & 0.18$^{+0.05}_{-0.07}$ & 12.3$^{+7.2}_{-2.2}$ & 8.1$^{+1.7}_{-2.6}$&19.6$^{+4.7}_{-1.3}$& $0.71^{+0.04}_{-0.04}$ & BBH\\
GWTC-2 & GW190828$\underline{\;\;}$063405 & 0.38$^{+0.10}_{-0.15}$ & 32.1$^{+5.8}_{-4.0}$ & 26.2$^{+4.6}_{-4.8}$ & 54.9$^{+7.2}_{-4.3}$ & 0.75$^{+0.06}_{-0.07}$ & BBH\\
GWTC-2 & GW190930$\underline{\;\;}$133541 & 0.15$^{+0.06}_{-0.06}$ & 12.3$^{+12.4}_{-2.3}$ & 7.8$^{+1.7}_{-3.3}$&19.4$^{+9.2}_{-1.5}$& $0.72^{+0.07}_{-0.06}$ & BBH\\
GWTC-3 & GW191103$\underline{\;\;}$012549    & 0.20$^{+0.09}_{-0.09}$& 11.8$^{+6.2}_{-2.2}$ &7.9$^{+1.7}_{-2.4}$ &19.0$^{+3.8}_{-1.7}$& 0.75$^{+0.06}_{-0.05}$ & BBH\\ 
GWTC-3 & GW191126$\underline{\;\;}$115259 & 0.30$^{+0.12}_{-0.13}$ & 12.1$^{+5.5}_{-2.2}$ & 8.3$^{+1.9}_{-2.4}$ & 19.6$^{+3.5}_{-2.0}$ & 0.75$^{+0.06}_{-0.05}$ & BBH\\
GWTC-3 & GW191127$\underline{\;\;}$050227 & 0.57$^{+0.40}_{-0.29}$ & 53$^{+47}_{-20}$ & 24$^{+17}_{-14}$&76$^{+39}_{-21}$& $0.75^{+0.13}_{-0.29}$ & BBH\\
GWTC-3 & GW191204$\underline{\;\;}$110529 & 0.34$^{+0.25}_{-0.18}$ & 27.3$^{+11.0}_{-6.0}$ & 19.3$^{+5.6}_{-6.0}$ & 45.0$^{+8.6}_{-7.6}$ & 0.71$^{+0.12}_{-0.11}$ & BBH\\
GWTC-3 & GW191204$\underline{\;\;}$171526 & 0.13$^{+0.04}_{-0.05}$ & 11.9$^{+3.3}_{-1.8}$ & 8.2$^{+1.4}_{-1.6}$&19.21$^{+1.79}_{-0.95}$& $0.73^{+0.03}_{-0.03}$ & BBH\\
GWTC-3 &  GW191219$\underline{\;\;}$163120$^{\dagger}$ & 0.11$^{+0.05}_{-0.03}$ & 31.1$^{+2.2}_{-2.8}$ & 1.17$^{+0.07}_{-0.06}$ & 32.2$^{+2.2}_{-2.7}$ & 0.14$^{+0.06}_{-0.06}$ & BH-NS\\
GWTC-3 & GW200105$\underline{\;\;}$162426$^\dagger$ & 0.06$^{+0.02}_{-0.02}$ & 9.0$^{+1.7}_{-1.7}$ & 1.91$^{+0.33}_{-0.24}$&10.7$^{+1.5}_{-1.4}$& $0.43^{+0.05}_{-0.02}$ & BH-NS\\
GWTC-3 & GW200112$\underline{\;\;}$155838 & 0.24$^{+0.07}_{-0.08}$ & 35.6$^{+6.7}_{-4.5}$ & 28.3$^{+4.4}_{-5.9}$ & 60.8$^{+5.3}_{-4.3}$ & 0.71$^{+0.06}_{-0.06}$ & BBH\\
GWTC-3 & GW200115$\underline{\;\;}$042309$^{\dagger}$ & 0.06$^{+0.03}_{-0.02}$ & 5.9$^{+2.0}_{-2.5}$ & 1.44$^{+0.85}_{-0.29}$&7.2$^{+1.8}_{-1.7}$& $0.42^{+0.09}_{-0.05}$ & BH-NS\\
GWTC-3 & GW200128$\underline{\;\;}$022011 & 0.56$^{+0.28}_{-0.28}$ & 42.2$^{+11.6}_{-8.1}$ & 32.6$^{+9.5}_{-9.2}$&71$^{+16}_{-11}$& $0.74^{+0.10}_{-0.10}$ & BBH\\
GWTC-3 & GW200129$\underline{\;\;}$065458 & 0.18$^{+0.05}_{-0.07}$ & 34.5$^{+9.9}_{-3.2}$ & 28.9$^{+3.4}_{-9.3}$&60.3$^{+4.0}_{-3.3}$& $0.73^{+0.06}_{-0.05}$ & BBH\\
GWTC-3 &  GW200208$\underline{\;\;}$222617 & 0.66$^{+0.54}_{-0.28}$ & 51$^{+104}_{-30}$ & 12.3$^{+9.0}_{-5.7}$ & 61$^{+100}_{-25}$ & 0.83$^{+0.14}_{-0.27}$ & BBH\\
GWTC-3 & GW200220$\underline{\;\;}$061928 & 0.90$^{+0.55}_{-0.40}$ & 87$^{+40}_{-23}$ & 61$^{+26}_{-25}$&141$^{+51}_{-31}$& $0.71^{+0.15}_{-0.17}$ & BBH\\
GWTC-3 & GW200224$\underline{\;\;}$222234 & 0.32$^{+0.08}_{-0.11}$ & 40.0$^{+6.9}_{-4.5}$ & 32.5$^{+5.0}_{-7.2}$ & 68.6$^{+6.6}_{-4.7}$ & 0.73$^{+0.07}_{-0.07}$ & BBH\\
GWTC-3 & GW200306$\underline{\;\;}$093714 & 0.38$^{+0.24}_{-0.18}$ & 28.3$^{+17.1}_{-7.7}$ & 14.8$^{+6.5}_{-6.4}$&41.7$^{+12.3}_{-6.9}$& $0.78^{+0.11}_{-0.26}$ & BBH\\
GWTC-3 & GW200308$\underline{\;\;}$173609 & 0.83$^{+0.32}_{-0.35}$ & 36.4$^{+11.2}_{-9.6}$ & 13.8$^{+7.2}_{-3.3}$&47.4$^{+11.1}_{-7.7}$& $0.91^{+0.03}_{-0.08}$ & BBH\\
GWTC-3 & GW200322$\underline{\;\;}$091133 & 0.60$^{+0.84}_{-0.30}$ & 34$^{+48}_{-18}$ & 14.0$^{+16.8}_{-8.7}$&53$^{+38}_{-26}$& $0.78^{+0.16}_{-0.17}$ & BBH\\
\hline
\hline
\end{tabular}
\end{table*}

\section{Number of high-energy events for BH-BH and BH-NS coalescences}

How many ultra-high energy events would be produced during binary coalescences? Since the acceleration of particles requires the presence of magnetic fields, in the BH-NS coalescence we can pinpoint that the neutron star provides intense magnetic fields onto the companion black hole. In addition, the presence of charges is guaranteed by the particles that extend around the neutron star following, e.g., a Goldreich-Julian density profile \cite{julian,mestel}. Throughout the coalescence, the star induces charges into the BH ISCO, producing a surface charge density profile of $q_{ISCO}/(2\pi r_{ISCO}^2)\sim \Omega B_0 R^3\delta r/(cr^ 3)$, where $\Omega$ is the angular velocity of the NS, $B_0$ the NS surface magnetic field, $R$ the radius of the star, $\delta r$ the distance of the surface of the NS to the ISCO of the BH and $r$ the distance of the center of the NS to the ISCO. To estimate upper limits in the number of events we will assume that the flow of charges through the ISCO is following geodesics of an extremal BH ($a \rightarrow 1$), for particles with maximal angular momenta \cite{banados}. This leads to the presence of collisional charges with speeds $v_{ISCO} \sim 0.3 c$, which produce a collisional electric current governed by the BSW geodesics. It is calculated by assuming a stream of charged particles moving inside the column connecting the neutron star and the ISCO of the black hole, i.e., $i \sim 2\pi R^2 \Omega B_0 v_*/c$, where $v_*$ is the velocity of the particles flowing from the NS. We take, as an example, a BH-NS binary with 10 $M_\odot$ BH and an NS with $R \sim 12$ km. Assuming $r \sim R + \delta r$, i.e., the NS is close to the ISCO, it follows that $v_* \rightarrow v_{ISCO}$. Therefore, the mean reaction time ($t_r$) for collisions of charged particles at the ISCO is $t_r \sim q_{ISCO}/i \sim r_{ISCO}^2\delta r/[v_{ISCO}R^2]$.
For $\delta r \sim (10^2-10^5)$ cm, for example, a conservative upper limit to $\dot{N}_{U.L.}$ would be $\sim 1/t_r \sim 10^4-10^7$ high energy events per second. During the merger time at which all neutron star charge content is absorbed, i.e., in a time $t_{merger}$ equal to $(0.01-2)$s (for the BH and NS masses assumed in the main text and GW frequencies around 100-400 Hz \citep{2018PhRvD..97b3016A}), a total number of high energy events $N = \dot{N}_{U.L.}t_{merger} = (10^2-10^7)$ is reached for this BH-NS coalescence.

Consider now the case of high-energy particles in BBH systems. The presence of magnetic fields and charged particles in the coalescence of BBHs would demand that at least one of the black holes should have an accretion disk. Assuming that the accretion column is passing through the BH ISCO, the mean reaction time for ion particles of the plasma can be estimated as follows. If the plasma has relativistic drift velocities $v_d$ at the accretion column, this would produce few high-energy collision events. However, close to the ISCO of the extremal Kerr BH, we should impose that the plasma particles will follow the geodesics of the strong gravitational field and, therefore, $v_d \rightarrow v_{ISCO} \sim 0.3 c$. Thus, the cross-section of two charged ions with charges $Z_1 e$ and $Z_2 e$ and masses $m_1$ and $m_2$, respectively, is inversely proportional to $v_d^4$ \cite{goldston1995,hazeltine2018}:
\[
\sigma \sim \left[\frac{Z_1 Z_2 e^2}{4\pi \epsilon_0}\right]^2 \frac{8\pi}{m_1 m_2 v_d^4} \ln \Lambda,
\]

\noindent where $\Lambda \sim \frac{4\pi \epsilon_0^{1.5} (kT)^{1.5}}{e^3 n_e^{0.5}}$ is the plasma parameter. For a $n_e \sim 10^{14}$cm$^{-3}$, $\ln \Lambda \sim 15$ plasma accreting a 10 $M_\odot$ BH, this will lead to $\sigma \sim  10^{-23}$ cm$^2$ for any proton-proton collision and $\sim 10^{-21}$ cm$^2$ for any iron-iron collision.  The mean reaction time in this case is $t_r \sim n_e^{-1} v_d^{-1} \sigma^{-1} \sim  1$ for proton collisions and $\sim 10^{-2}$ for iron collisions, i.e., the upper limit to $\dot{N}_{U.L.}$ is $\sim 1-10^2$ high energy events per second. During the merger time at which all charge content of the accretion disk is absorbed, which in our case is around $(0.01-2)$ s, the upper limit to the total number of high energy events is $N = \dot{N}_{U.L.}t_{merger} = (10^{-2}-10^2)$. However, accretion disks of low mass BHs could have larger values of $n_e$, for instance, $10^{17}-10^{20}$ cm$^{-3}$ \citep{2016MNRAS.462..751G}. In this case, the number of high-energy events before the merger could be increased by a factor of $10^3-10^6$ with respect to those coming from a conservative $n_e$. Therefore, for BBH and BH-NS mergers, up to millions of ultra-high energy events would be expected. 

\bibliography{example}

\end{document}